\documentclass[%
 aip,
 jmp,%
 amsmath,amssymb,
 reprint,%
]{revtex4-1}

\usepackage{graphicx}
\usepackage{dcolumn}
\usepackage{bm}
\usepackage{amsmath,amssymb}
\usepackage{graphicx,float}
\usepackage{times,txfonts}
\usepackage{color}
\usepackage{multirow}
\usepackage{bbold}
\usepackage{color}
\usepackage{soul}
\usepackage{hyperref}
\usepackage{natbib}
\usepackage{nicefrac}

\newcommand{\abs}[1]{\left| #1 \right|}
\newcommand{\bra}[1]{\langle #1|}
\newcommand{\ket}[1]{\vert#1\rangle}
\newcommand{\braket}[2]{\left\langle {#1{\left| \vphantom{#1 #2} \right.} #2} \right\rangle}

\renewcommand{\epsilon}{\varepsilon}

\definecolor{hugoColor}{RGB}{59,134,255}
\definecolor{YellowOrange}{RGB}{226,154,2}
\sethlcolor{YellowOrange}
\sethlcolor{lightblue}

\DeclareMathOperator{\Ai}{Ai}
\DeclareMathOperator{\sign}{sign}
\begin{document}


\title[Structured Quantum Projectiles]{Structured Quantum Projectiles}

\author{Hugo~Larocque}
\affiliation{Department of Physics, University of Ottawa, 25 Templeton St., Ottawa, Ontario, K1N 6N5 Canada}
%
\author{Robert~Fickler}
\email{robertfickler@web.de}
\affiliation{Department of Physics, University of Ottawa, 25 Templeton St., Ottawa, Ontario, K1N 6N5 Canada}
\affiliation{current address: Institute for Quantum Optics and Quantum Information (IQOQI), Austrian Academy of Sciences, Boltzmanngasse 3, A-1090 Vienna, Austria}
\author{Eliahu~Cohen}
\affiliation{Department of Physics, University of Ottawa, 25 Templeton St., Ottawa, Ontario, K1N 6N5 Canada}
\affiliation{Faculty of Engineering and the Institute of Nanotechnology and Advanced
Materials, Bar Ilan University, Ramat Gan 5290002, Israel}
\author{Vincenzo~Grillo}
\affiliation{CNR-Istituto Nanoscienze, Centro S3, Via G Campi 213/a, I-41125 Modena, Italy}
\author{Rafal~E.~Dunin-Borkowski}
\affiliation{Ernst Ruska-Centre for Microscopy and Spectroscopy with Electrons and Peter Gr\"unberg Institute, Forschungszentrum J\"ulich, 52425 J\"ulich, Germany}
\author{Gerd Leuchs}
\affiliation{Department of Physics, University of Ottawa, 25 Templeton St., Ottawa, Ontario, K1N 6N5 Canada}
\affiliation{Max-Planck-Institut f\"ur die Physik des Lichts, Staudtstra{\ss}e 2, 91058 Erlangen, Germany}
\affiliation{Institut f\"ur Optik, Information und Photonik, Universit\"at Erlangen-N\"urnberg, Staudtstra{\ss}e 7/B2, 91058 Erlangen, Germany}
\author{Ebrahim~Karimi}
\affiliation{Department of Physics, University of Ottawa, 25 Templeton St., Ottawa, Ontario, K1N 6N5 Canada}
\affiliation{Max-Planck-Institut f\"ur die Physik des Lichts, Staudtstra{\ss}e 2, 91058 Erlangen, Germany}

\date{\today}

\begin{abstract}
Matter wave interferometry is becoming an increasingly important technique in quantum metrology. However, unlike its photonic counterpart, this technique relies on the interference of particles possessing a non-zero rest mass and an electric charge. Matter waves, thus, can experience alterations in their wave-like features while propagating through uniform fields to which a linear potential can be attributed. Here, we derive analytical expressions for \emph{structured} matter waves subjected to linear potentials. We show that the center of mass of corresponding to these wavefunctions follows the classical parabolic trajectory attributed to this potential and also provide the additional phase profile acquired by the wave upon propagation. Furthermore, we find that these features are identical for any structured wave, thus significantly simplifying the action of quantum effects pertaining to this potential in applications relying on structured quantum waves.
\end{abstract}

\maketitle

\section{\label{sec:level1}Introduction}
The wave nature of massive quantum particles is one of the most prominent paradigms of quantum physics. On one hand, quantum phenomena such as superposition and interference of single massive particles such as electrons~\cite{Has09}, neutrons~\cite{Rau00}, atoms~\cite{Cro09} and even molecules~\cite{Hor12} are used to study fundamental questions of physics \cite{Arn14}. On the other hand, matter wave interferometry has become a powerful tool for advanced quantum technologies in information science and high precision metrology tasks\cite{lucke2011}.  Adapting ideas and techniques from structured photonics~\cite{Har15,Rub16}, shaping the transverse wavefronts of matter waves, which is directly associated to the wave nature of quantum objects, has attracted much attention recently~\cite{Bli17,Llo17,Lar18}. Particularly interesting examples of such structured quantum waves are those with twisted wavefronts, i.e. with azimuthally varying phase profiles, as this leads to a quantized orbital angular momentum (OAM) carried by these freely propagating quantum particles~\cite{All92}. For instance, early realizations of such waves consisted of photo-electrons emitted in multi-photon ionization of atoms. Such electrons were shown to exhibit highly anisotropic angular distributions as a result of the high OAM of the various outgoing partial electron waves.~\cite{Smith88}
It was already pointed out in an early theoretical proposal~\cite{Bli07} that for charged matter waves, e.g. electrons, this twisted structure will lead to an additional unbounded magnetic dipole moment. Following this initial theoretical discussion, various experimental studies have confirmed this prediction. Moreover, twisted electrons have not only been used to perform fundamental tests but also to equip modern electron microscopes with a novel type of magnetic nano-sensor. In addition to structured electrons, which can be considered an already well-established field, the ability of twisting the wavefronts of heavier quantum systems such as neutrons has also been investigated~\cite{Cla15,Sar:16}. Interestingly, twisting the wavefront of neutron beams could enable a novel way to study the internal structure of neutrons and as such opens a new field of applications of more massive structured quantum waves~\cite{Lar18Neutron}. In addition to their internal structures, large freely propagating systems like neutrons, atoms and molecules are also strongly affected by the influence of gravity~\cite{Zim18,Sev17}. Hence, a natural open question to ask oneself is to what extent a linear potential, such as the gravitational field, might affect the shape and trajectory of the structured matter wave.

In this article, we investigate this question by analytically solving the paraxial wave equation for higher order Gaussian matter waves in linear potentials such as a gravitational potential. We find that the center of mass for Hermite-Gaussian wavefunctions, as well as OAM-carrying Laguerre-Gaussian wavefunctions, follows the classical, parabolic trajectory of a falling particle, irrespectively of their mode order or OAM value. We further show that they only accumulate a cubic phase term upon propagation, which indicates that the special features of structured matter waves, such as the OAM of twisted matter waves, should persist even for very heavy quantum systems such as large molecules. Finally, we briefly discuss the effect of other linear potentials arising e.g. from constant external electric fields, and the implications of our results on current research efforts. For instance, a linear electric potential acting on charged particles can in principle be used to simulate and predict the effect of gravitational fields on twisted matter waves.

\section{Schr\"odinger equation for a massive particle in the presence of a linear potential}

The wavefunction $\Psi(\mathbf{r},t)$ of a non-relativistic particle experiencing a potential of the form $V(\mathbf{r})=\alpha x$, where $\mathbf{r} = x \,\mathbf{e}_x+y \,\mathbf{e}_y+z \,\mathbf{e}_z$ is the position vector expressed in Cartesian coordinates and $\alpha$ is a constant, satisfies the Schr\"odinger equation
\begin{equation}
	\label{eq:Seq}
	\left(-\frac{\hbar^2}{2m}\boldsymbol{\nabla}^2 + \alpha x \right)\Psi(\mathbf{r},t)  = i\hbar\, \partial_t\Psi(\mathbf{r},t).
\end{equation}
Here, $\hbar$ is the reduced Planck constant, $m$ is the particle's mass, $\partial_t$ is the time partial derivative, and $\boldsymbol{\nabla}^2$ is the Laplacian. 
\begin{figure}[t]
	\centering
	\includegraphics[width=12cm]{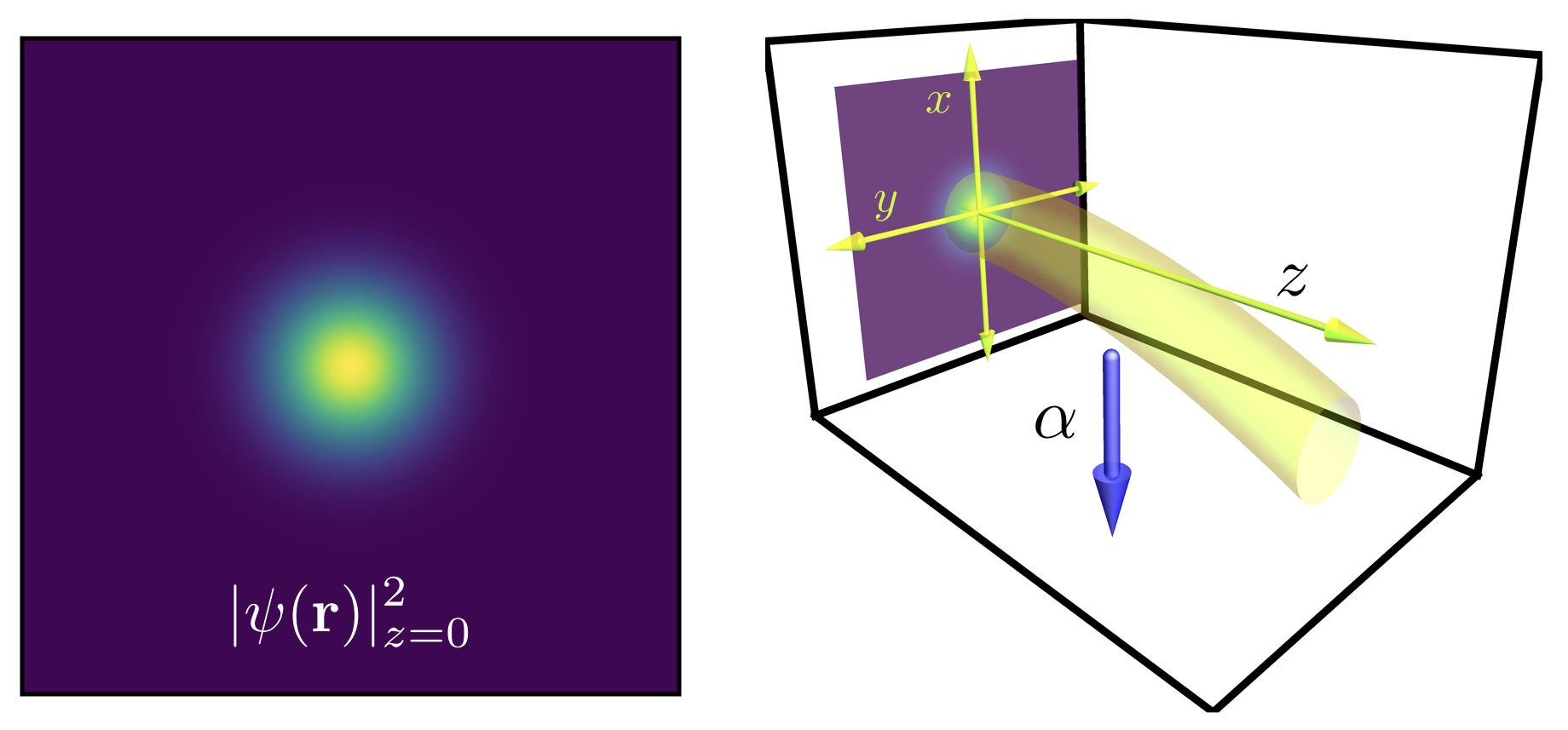}
	\caption{{\bf Schematic diagram of a matter wave propagating within a linear potential}. A matter wave defined by a known wavefunction at the $z=0$ plane propagates along the $z$ axis under the influence of a linear potential of strength $\alpha$ acting along the $x$ direction.}
	\label{fig:schematic}
\end{figure}
As outlined in Fig.~\ref{fig:schematic}, our goal is to find solutions to Eq.~(\ref{eq:Seq}) whose transverse formulation is known for a given longitudinal coordinate, in our case $z=0$, and examine how they evolve upon propagation due to the presence of the linear potential. We may assume that our wavefunction possesses a well-defined central energy $\mathcal{E}_0$ and longitudinal momentum $p_0$ such that $\mathcal{E}_0\approx p_0^2/2m$. This allows us to express our wavefunction as $\Psi(\mathbf{r},t)=\psi(\mathbf{r})\exp{(i(p_0 z- \mathcal{E}_0 t)/\hbar)}$. This form, along with the assumptions that $\psi(\mathbf{r})$ is slowly varying and that the transverse extent of the wavefunction is much larger than the particle's de Broglie wavelength, allows the Schr\"odinger equation to be reduced to the paraxial wave equation~\cite{Bli07} in the presence of a potential, i.e.
\begin{equation}
	\left(-\frac{\hbar^2}{2m}\boldsymbol{\nabla}_\perp^2 + \alpha x\right) \psi(\mathbf{r})=i\hbar \left(\frac{p_0}{m}\right)\,\partial_z\psi(\mathbf{r}),
\end{equation}
where $\boldsymbol{\nabla}_\perp^2=\partial^2_x+\partial^2_y$ is the transverse Laplacian. Assuming that the wavefunction is known at a certain transverse plane (assigned for convenience to $z=0$) $\ket{\psi}_0$, then it can be found at any $z$ plane $\ket{\psi}_z$ by means of the propagation operator $\hat{U}$ attributed to the above paraxial wave equation. Namely, 
\begin{equation}\label{eq:Uevolution}
	\ket{\psi}_z =\hat{U}(z)\,\ket{\psi}_0\,,
\end{equation}
where 
\begin{eqnarray}\nonumber
	\hat{U}(z)&=&\exp{\left(-i\frac{m z}{\hbar p_0}\,\hat{\mathcal{H}}_0\right)},\\
	\hat{\mathcal{H}}_0&=&-\frac{\hbar^2}{2m}\boldsymbol{\nabla}_\perp^2+ \alpha \hat{x}.
\end{eqnarray}
The position representation of the wavefunction $\psi(\mathbf{r}_\perp,z)=\braket{\mathbf{r}_\perp,z}{\psi}$ can then be explicitly calculated with the completeness relation $\int \ket{\mathbf{r}_\perp}\bra{\mathbf{r}_\perp}\,d^2 \mathbf{r}_\perp =1$
\begin{equation}
\label{eq:propagation}
	\psi\left(\mathbf{r}_\perp,z\right) = \int d^2 \mathbf{r}'_\perp K\left(\mathbf{r}_\perp,z;\mathbf{r}'_\perp,0\right) \, \psi\left(\mathbf{r}'_\perp,0\right),
\end{equation}
where $K\left(\mathbf{r}_\perp,z;\mathbf{r}'_\perp,0\right)=\bra{\mathbf{r}_\perp} \hat{U}(z) \,\lvert{\mathbf{r}'_\perp}\rangle$ consists of the physical system's propagation kernel and links the wavefunction at a certain propagation distance $z$ with its initial formulation in the $z=0$ plane. An expression for the propagation kernel can readily be derived provided that we know the eigenstates of $\hat{\mathcal{H}}_0$, that is the eigenvectors $\ket{n}$ that satisfy the eigenvalue equation $\hat{\mathcal{H}}_0 \ket{n} =\epsilon_n \ket{n}$. Using separation of variables, i.e. $\braket{\mathbf{r}_\perp}{n}=\chi(x)\,\phi(y)$, one may divide our initial eigenvalue equation into the following ones
\begin{eqnarray}\label{eq:eigenequation}
	\left(-\frac{\hbar^2}{2m}\,\frac{d^2}{dx^2} + \alpha x\right) \chi(x) = \epsilon_x \, \chi(x), \\
	\left(-\frac{\hbar^2}{2m}\,\frac{d^2}{dy^2}\right) \phi(y)= \epsilon_y \, \phi(y),
\end{eqnarray}
which respectively yield Airy and plane wave solutions, i.e.,
\begin{eqnarray}\label{eq:x-eigenstate}
	\chi(x) &=& \frac{\abs{\tau^{1/3}}}{\abs{\alpha}^{1/2}}\Ai\left(\tau^{1/3}\left(x-\frac{\epsilon_x}{\alpha}\right)\right)\\ \label{eq:y-eigenstate}
	\phi(y) &=& \frac{1}{\sqrt{2\pi}}\exp(\pm i k_y y)
\end{eqnarray}
where $k_y=\sqrt{2m\epsilon_y}/\hbar$ and $\tau= {2m\alpha}/{\hbar^2}$. $\chi(x)$ and $\phi(y)$ are normalized such that $\langle{\chi_{\epsilon^1_{x}}}\vert\,{\chi_{\epsilon^2_{x}}}\rangle =\delta(\epsilon^2_{x}-\epsilon^1_{x})$ and $\langle{\phi_{k^1_y}}\vert\,{\phi_{k^2_y}}\rangle =\delta(k^2_{y}-k^1_{y})$, where $\delta$ is the Dirac delta function, and form complete bases. Note that the eigen-equation for $\chi(x)$, Eq.~(\ref{eq:eigenequation}), can also admit solutions in the form of Airy functions of the second kind $\text{Bi}(x)$. However, given that these functions diverge as $x \rightarrow +\infty$, then they do not satisfy our system's boundary conditions and are thereby excluded from our solution (see \cite{Bal79,VB13,Zim18} for further discussions on the Airy function).

\section{Kernel of propagation and propagated wavefunctions}
With the eigenstates of Eqs.~(\ref{eq:x-eigenstate},\ref{eq:y-eigenstate}), we may rewrite our kernel as
\begin{eqnarray}\nonumber
	K(\mathbf{r}_\perp,z;\mathbf{r}'_\perp,0) &=& \int d\epsilon_x\,d k_y \,\bra{\mathbf{r}_\perp} \hat{U}(z)\,\lvert{n(\epsilon_x,k_y)}\rangle \langle{n(\epsilon_x,k_y)} \lvert{\mathbf{r}'_\perp}\rangle \\
	&=& K(x,z;x',0)\, K(y,z;y',0),
\end{eqnarray}
where we used the separable nature of our eigenstates to split our propagation kernel into two components
assigned to different coordinates. These components are given by
\begin{eqnarray}
	K(x,z;x',0)&=&\frac{\abs{\tau^{2/3}}}{\abs{\alpha}}\int d\epsilon_x \exp{\left(-i\left(\frac{m z}{\hbar p_0}\right)\epsilon_x\right)}\cr\cr
	&\times&\Ai\left(\tau^{1/3}\left(x-\frac{\epsilon_x}{\alpha}\right)\right)\,\Ai\left(\tau^{1/3}\left(x'-\frac{\epsilon_x}{\alpha}\right)\right)\\
	K(y,z;y',0)&=&\frac{1}{2\pi}\int dk_y \exp{\left(-i\left(\frac{\hbar z}{2p_0}\right)k_y^2\right)}\exp\left(i k_y(y-y')\right).\cr\nonumber
\end{eqnarray}
The integral form of $K(y,z;y',0)$ simply consists of a Gaussian integral and can be readily evaluated, i.e.,
\begin{equation}
	\label{eq:PropY}
	K(y,z;y',0) = \sqrt{\frac{p_0}{ i 2\pi\hbar z}} \exp{\left(i \frac{p_0}{2\hbar z}\left(y-y'\right)^2\right)}.
\end{equation}
As for $K(x,z;x',0)$, it can be evaluated by making use of the following integral~\cite{Val10},
\begin{eqnarray}
	\frac{1}{\lvert\alpha\beta\rvert} \int&du& \, e^{i\lambda u} \Ai\left(\frac{u+a}{\alpha}\right) \Ai\left(\frac{u+b}{\beta}\right) \\
	&=&\frac{1}{2\sqrt{\pi}\abs{\lambda}^{1/2}\abs{\alpha}^{3/2}} \exp\left(-i f\left(\alpha,\beta,\lambda,a,b\right)\right),
\end{eqnarray}
where
\begin{eqnarray}\nonumber
	f\left(\alpha,\beta,\lambda,a,b\right)= \frac{\alpha^3\lambda^3}{12}- \frac{(a-b)^2}{4 \alpha^3 \lambda} +\frac{\lambda(a+b)}{2} +\frac{\pi}{4}\sign(\alpha \lambda)
\end{eqnarray}
which holds when $\alpha = \beta$. By using this result, we obtain the following Kernel
\begin{eqnarray}
	\label{eq:PropX}
	K(x,z;x',0)= \sqrt{\frac{p_0}{ i 2\pi\hbar z}}  \exp\left(-i f\left(\alpha,p_0,z,x,x'\right)\right),
\end{eqnarray}
where
\begin{eqnarray}\nonumber
	f\left(\alpha,p_0,z,x,x'\right)= \frac{\alpha^2 m^2 z^3}{24\, \hbar p_0^3}- \frac{1}{2}\frac{p_0}{\hbar z}\left(x-x'\right)^2 + \frac{\alpha m z}{2\hbar p_0}\left(x+x'\right).
\end{eqnarray}
The propagators in Eq.~(\ref{eq:PropX}) and Eq.~(\ref{eq:PropY}) may then be used in conjunction with Eq.~(\ref{eq:propagation}) to calculate the evolution of a Gaussian matter wave upon propagation through a linear potential. Unlike plane waves, Gaussian waves are defined by a finite transverse extent parametrized by a quantity $w_0$ known as their waist. This modulation accounts for several physical traits attributed to matter waves in general, such as their finite energy and their broadening upon propagation due to diffraction~\cite{Lar18}. To generalize this concept for the case of structured matter waves, i.e. matter waves that can be expressed as superpositions of higher-order Gaussian modes, we will consider initial wavefunctions of the form
\begin{equation}
	\psi(\mathbf{r}'_\perp,0) \propto \exp\left(-\left(\frac{x^2+y^2}{w_0^2}\right)\right) H_m\left(\frac{\sqrt{2}x}{w_0}\right)H_n\left(\frac{\sqrt{2}y}{w_0}\right),
\end{equation}
where $w_0$ is the beam's waist while $H_m(.)$ is the $m^\text{th}$ order Hermite polynomial~\cite{Sie:86}. Wavefunctions of this form are known as \textit{Hermite-Gauss} wavefunctions, and provide a basis in Cartesian coordinates in which any arbitrary wavefunction satisfying the paraxial wave equation can be decomposed. With this initial wavefunction along with the previously derived propagation kernels, we obtain the following $x$ and $y$ components of the wavefunction
\begin{eqnarray}\label{eq:wavefunctionX}\nonumber
	\psi_x(x,z) &\propto& \sqrt{\frac{1}{1+i\zeta}}\left(\frac{1-i\zeta}{1+i\zeta}\right)^{m/2}\exp\left(-i \left( \frac{1}{6}\frac{\alpha^2 m^2 z^3}{\hbar p_0^3}\right)\right)\\ \nonumber
	&\times&\exp\left(-i\frac{\alpha m z}{\hbar p_0}x\right)\exp\left(-\left(\frac{\left(x+\frac{m\alpha z^2}{2p_0^2}\right)^2}{w_0^2(1+i\zeta)}\right)\right) \\ &&H_m \left(\frac{\sqrt{2} \, \left(x+\frac{m\alpha z^2}{2p_0^2}\right)}{w_0(1+\zeta^2)^{1/2}}\right),
\end{eqnarray}
\begin{eqnarray}\label{eq:wavefunctionY}
	\psi_y(y,z) &\propto&  \sqrt{\frac{1}{1+i\zeta}} \left(\frac{1-i\zeta}{1+i\zeta}\right)^{n/2}\\ \nonumber
	&\times&\exp\left(-\left(\frac{y^2}{w_0^2(1+i\zeta)}\right)\right) H_n \left(\frac{\sqrt{2} \,
y}{w_0(1+\zeta^2)^{1/2}}\right),
\end{eqnarray}
where we defined $\psi_x$ and $\psi_y$ such that $\psi(x,y,z)=\psi_x(x,z)\,\psi_y(y,z)$, $\zeta=z/z_R$, and $z_R =(p_0 w_0^2)/(2\hbar)$ is the matter wave's Rayleigh range. In essence, the latter quantity consists of a longitudinal distance over which the wave does not experience significant diffraction~\cite{Lar18}.
\begin{figure}[h]
	\centering
	\includegraphics[width=12cm]{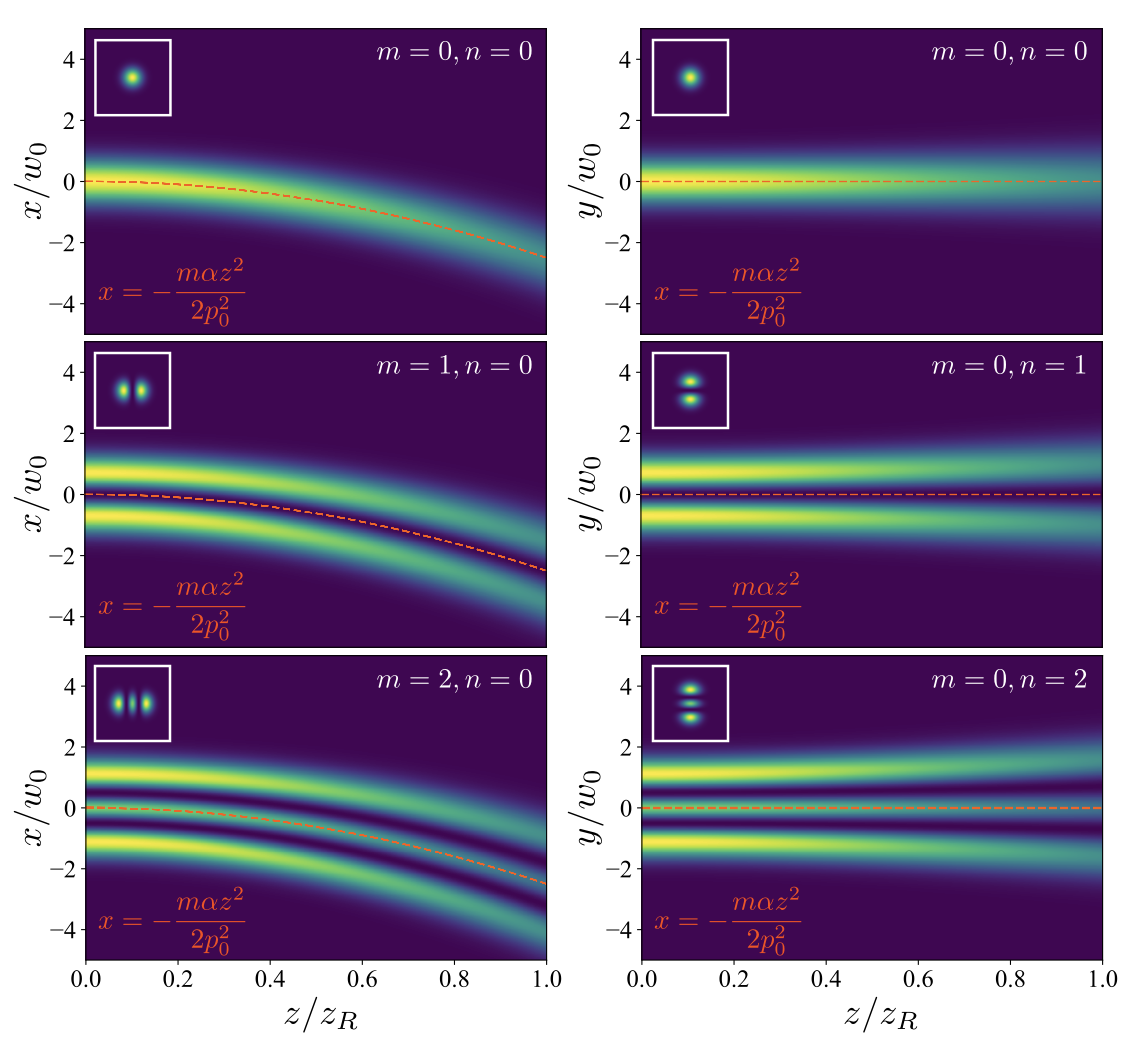}
	\caption{{\bf Propagation dynamics of Hermite-Gaussian wavefunctions}. Probability densities of Hermite-Gauss wavefunctions defined by different $m$ and $n$ indices upon propagation through a linear potential of strength $\alpha$. In addition to their modal indices, the wavefunctions are defined by a waist of $w_0$ and a longitudinal momentum of $p_0$. The beams are also plotted along the centroid formed by their probability densities, which is denoted as a red dotted line. The initial transverse profiles of these beams are shown as insets in their respective propagation plots.}
	\label{fig:HG}
\end{figure}
Several conclusions can be established from the above results. To begin with, the $y$ component of the wavefunction does not experience any alterations caused by the linear potential. As expected, its evolution is simply attributed to that of a free matter wave. As for the $x$ component of the wavefuncion, we can see that the linear potential affects a few of its features. First, the wavefunction acquires a $T^3$ phase upon propagation, i.e., a phase associated with the presence of a linear potential which increases cubically with a coordinate, $z$ in this case, attributed to the evolution of the wavefunction. Second, its probability density distribution $\abs{\psi_x}^2$ is centered along the classical trajectory attributed to a particle propagating in the presence of a constant force, i.e. $x(z) = x_0-({m\alpha z^2})/({2p_0^2})$, where $x_0$ is the particle's initial position along the $x$-axis. Finally, the second exponential term of Eq.~(\ref{eq:wavefunctionX}) adds a $z$-dependent momentum in the $x$-direction, thereby affecting the phase curvature of the wave upon propagation. Other than those attributes, the matter wave still experiences the same alterations upon propagation as those of a free matter wave. More specifically, its probability density, though shifted, preserves its shape upon propagation and diffracts at the same rate as
would a wave undergoing free propagation. These features can be readily observed in Fig.~\ref{fig:HG}, where the
probability density of the first so-called Hermite-Gaussian (HG) wavefunctions are plotted along the densities'
centroid.
\begin{figure}[t]
	\centering
	\includegraphics[width=11cm]{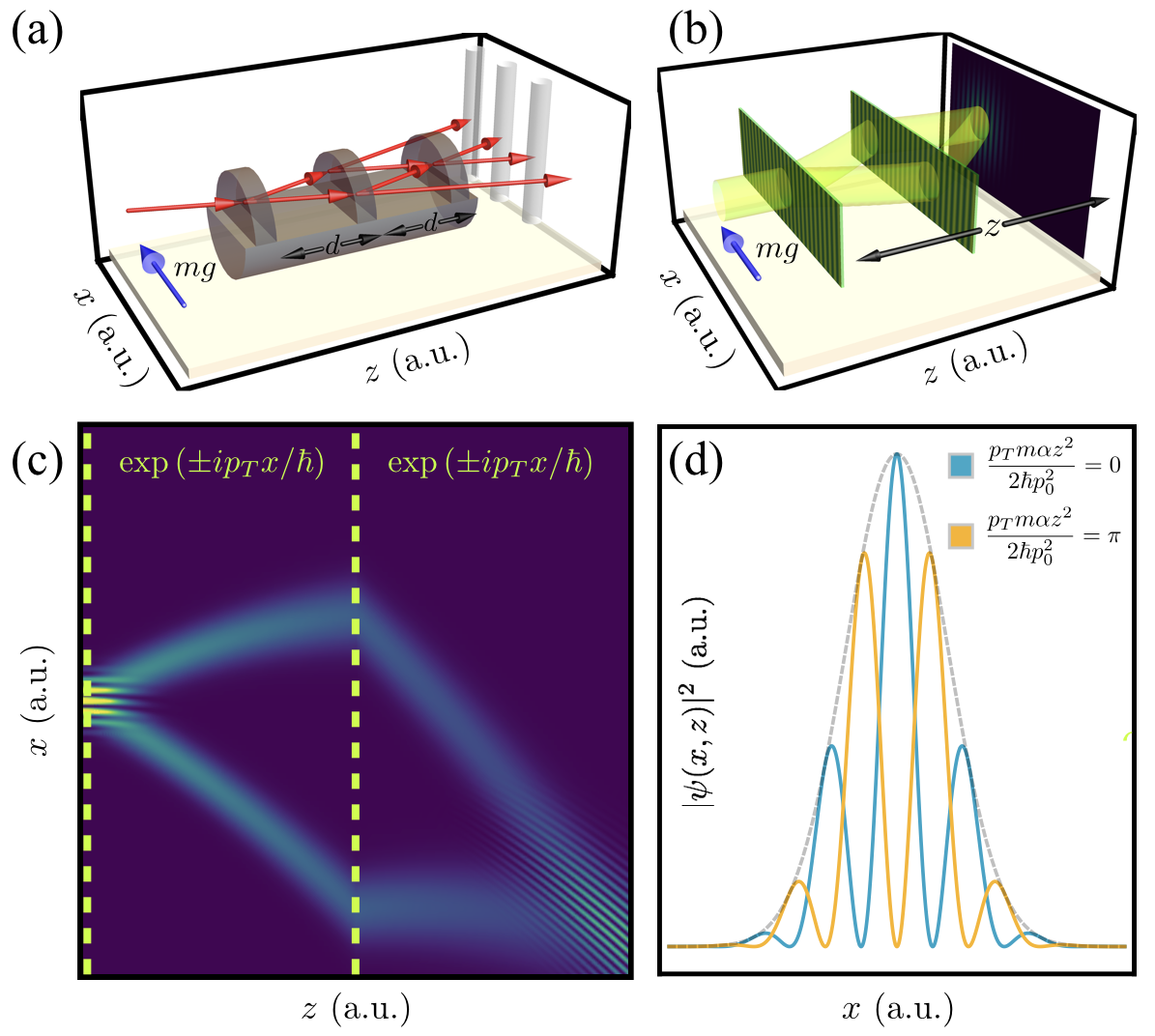}
	\caption{{\bf Interferometry with Gaussian wavefunctions experiencing a linear potential}. {(a)} Conventional Bragg grating interferometry apparatus for matter waves in a linear potential. {(b)} Apparatus replicating the interference effect seen in {(a)} by simulating the action of the gratings by phase elements that add a phase of $\exp{(i p_T x/\hbar)}$ to the matter wave. {(c)} Analytical probability density function of the wavefunction going through the apparatus shown in {(b)}. The positions of the phase elements are denoted by dotted lines. {(d)} Phase shift observed in the wavefunction upon varying the quantity $(p_T m\alpha z^2)/(2 \hbar p_0^2)$. The position of the wave is centered with respect to its center of mass.}
	\label{fig:interferometer}
\end{figure}

The veracity and practical importance of the above results can readily be attested by using the propagation Kernel in Eq.~(\ref{eq:PropX}) to derive the wavefunction of a matter wave experiencing a linear potential inside an interferometry experiment. An example of such interferometry is depicted in Fig.~\ref{fig:interferometer}{(a)}, where two Bragg gratings are used to separate and eventually recombine the matter wave with itself after being affected by a potential of $\alpha x=mgx$ -- where $g\simeq9.8 m/s^2$ is the gravitational acceleration constant. The phase shift observed at the output of such an apparatus is known to be $\Delta\Phi=4\pi\lambda g h^{-2}m^2d(d+a\cos{\theta})\tan{\theta}\sin{\phi}$, where $\lambda$ is the matter wave's de Broglie wavelength, $h$ is the Planck constant, $d$ is the distance separating the gratings, $a$ is the thickness of the gratings, $\theta$ is the gratings' Bragg angle, and $\phi$ is the angle between the direction along which the potential varies and the one along which the matter waves are diffracted~\cite{Col75,Abe08,Abe12}. As shown in Fig.~\ref{fig:interferometer}{(b)}, to simulate this experiment, we calculate the wavefunction of a matter wave in a linear potential propagating along a distance $z$ over which it is affected by two phase elements that add a phase of $\pm p_T x/\hbar$ to the wavefunction. The resulting probability density function is shown in Fig.~\ref{fig:interferometer}{(c)}, and the corresponding recombined wavefunction is observed to experience a phase shift of $\Delta\Phi= (p_T m\alpha z^2)/(2 \hbar p_0^2)$ as shown in Fig.~\ref{fig:interferometer}{(d)}. This expression is in agreement with the one used in Fig.~\ref{fig:interferometer}{(a)} given that $p_T = p_0 \tan{\theta}$, $a=0$, and $z =2d$.

\section{Expansion Coefficients}

\begin{figure}[t]
	\centering
	\includegraphics[width=11cm]{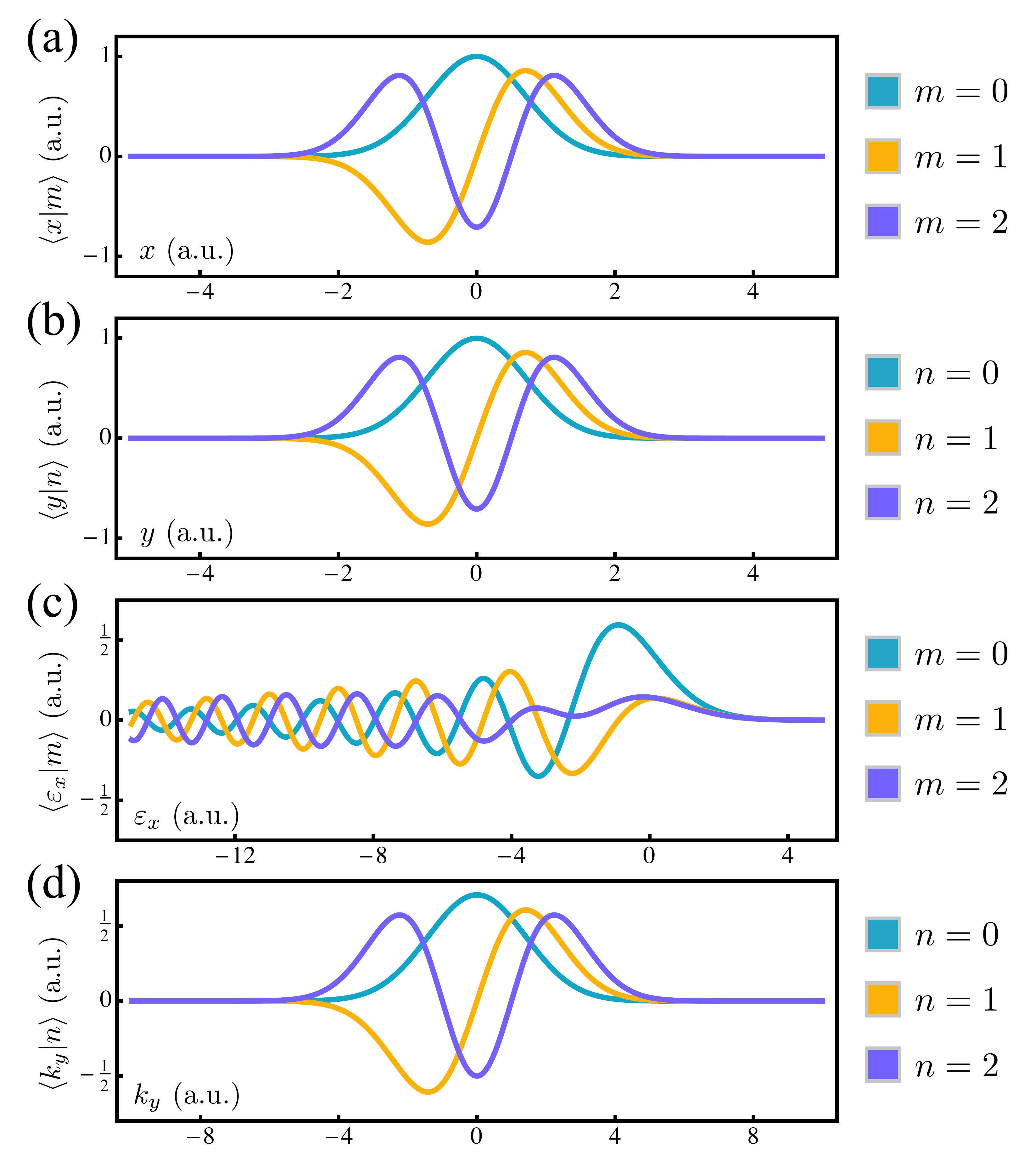}
	\caption{{\bf Eigenstate decomposition of Hermite-Gauss modes}. Wavefunctions $\langle{x,y}\rvert{m,n}\rangle=\langle{x}\rvert{m}\rangle\langle{y}\rvert{n}\rangle$ of the first Hermite-Gauss modes and their corresponding eigenstate expansion coefficients $\langle{\varepsilon_x,k_y}\rvert{m,n}\rangle=\langle{\varepsilon_x}\rvert{m}\rangle\langle{k_y}\rvert{n} \rangle$. Each function is scaled by a factor of $1/(2^{n/2}\sqrt{n!})$ or $1/(2^{m/2}\sqrt{m!})$.}
	\label{fig:expansion}
\end{figure}

In free-space, the spatial wavefunction of higher-order Gaussian modes also corresponds to its expansion coefficients in terms of transverse momenta components. In the presence of a linear potential, however, this correspondence breaks down given that the $x$ component of the system's eigenfunctions is expressed in terms of Airy functions as opposed to plane waves. Therefore, for an eigenstate $\ket{\psi}=\ket{\,m,n}$, where $\braket{\mathbf{r}}{\psi}$ is given by the product of the wavefunctions in Eqs.~(\ref{eq:wavefunctionX},\ref{eq:wavefunctionY}), the expansion coefficients $\langle{\varepsilon_x,k_y}\rvert{m,n}\rangle$=$\langle{\varepsilon_x}\rvert{m}\rangle\langle{k_y}\rvert{n}\rangle$ now involve integrations over Airy functions. Namely, whereas
$\langle{k_y}\rvert{n}\rangle$ still only involves a well-known Fourier transform, $\langle{\varepsilon_x}\rvert{m}\rangle$ is now defined as
\begin{equation}
	\langle{\varepsilon_x}\rvert{m}\rangle \propto
	\int_{-\infty}^{\infty}\Ai\left(\tau^{1/3}\left(x-\frac{\epsilon_x}{\alpha}\right)\right)e^{-{x^2}/{w_0^2}} H_m\left(\frac{\sqrt{2}x}{w_0}\right) \,dx
\end{equation}
which formally consists of the state's Airy transform that, for a function $f(x)$, is defined as~\cite{Val10,Wid79} $$\varphi_\alpha(y)=\frac{1}{\abs{\alpha}} \int_{-\infty}^\infty\Ai\left(\frac{y-x}{\alpha}\right)\,f(x)\,dx.$$ To solve the above integral, one can make use of the Airy transform of a Gaussian function which is given by~\cite{Val10}:
\begin{equation}	
	\frac{\sqrt{\pi}}{\abs{\alpha}}\exp{\left(\frac{1}{4\alpha^3}\left(y+\frac{1}{24\alpha^3}\right)\right)}\Ai\left(\frac{y}{\alpha}+\frac{1}{16\alpha^4}\right).
\end{equation}
With this relation and the generating function of the Hermite polynomials, the Airy transform of the function
$\exp(-x^2)H_m(\sqrt{2}x)$ can be derived as:
\begin{eqnarray}
	\label{eq:AiryHermite}
	\varphi_\alpha^{\text{HG}}(y) &\propto& \frac{\sqrt{\pi}}{\abs{\alpha}}\exp{\left(\frac{1}{4\alpha^3}\left(y+\frac{1}{24\alpha^3}\right)\right)} \\ \nonumber
	&&\sum_{n=0}^{m} {m \choose n}\, H_n\left(\frac{\sqrt{2}i}{8\alpha^3}\right)i^n \left(\partial_t^{m-n} \Ai{\left(\frac{y-\sqrt{2}t}{\alpha}+\frac{1}{16 \alpha^4}\right)}\right)\biggr|_{t=0}.
\end{eqnarray}
With the above equation, the expansion coefficients $\langle{\varepsilon_x,k_y}\rvert{m,n}\rangle$ may then be calculated analytically as shown in Fig.~\ref{fig:expansion}. These coefficients may be of use in several types of quantum mechanical calculations, such as those involving perturbation theory, that involve the derived accelerated structured matter waves. Furthermore, they can also be used to relate the propagated wavefunctions in Eq.~(\ref{eq:wavefunctionX}) to their Airy transforms, which we can expect to be given by the result extracted from Eq.~(\ref{eq:AiryHermite}) multiplied by an $\exp{(-i(mz/\hbar p_0)\varepsilon_x)}$ term.

\section{Vortex Dynamics}
\begin{figure}[t]
	\centering
	\includegraphics[width=11cm]{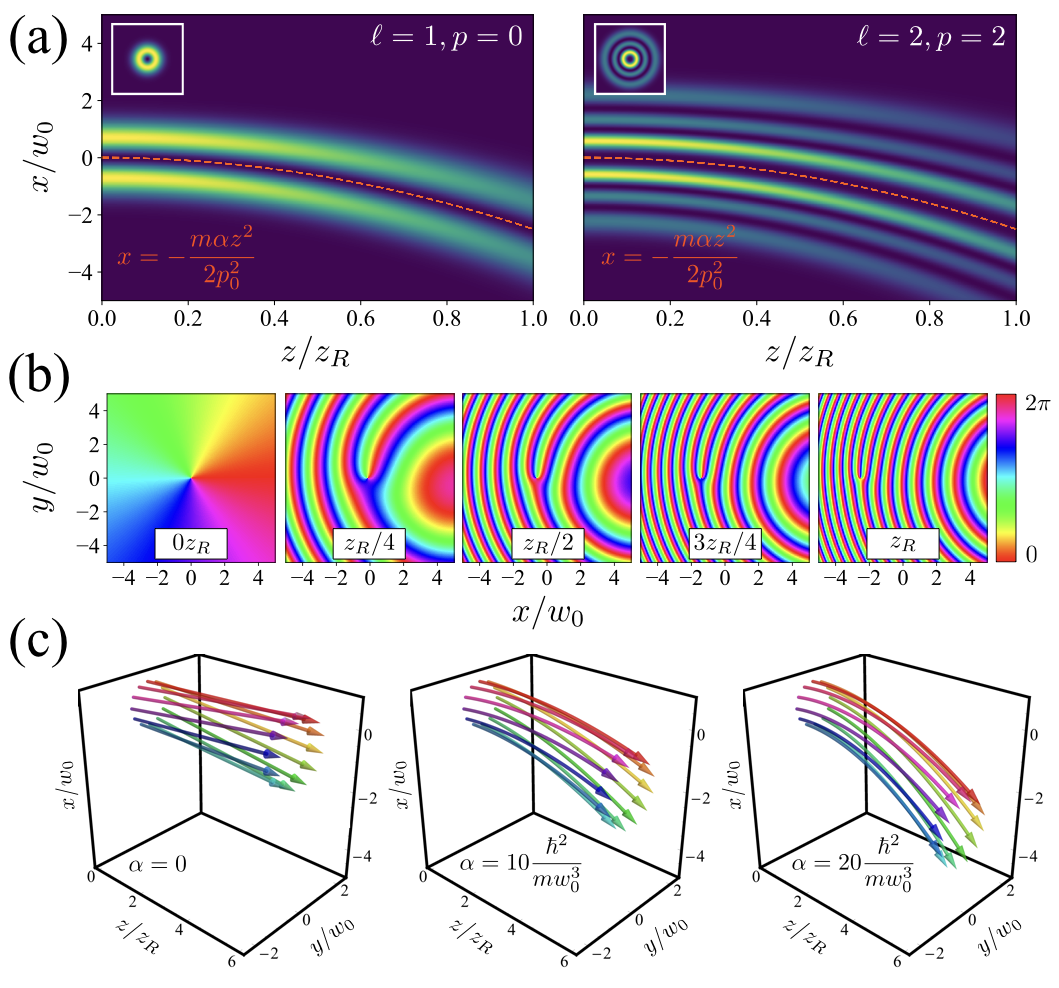}
	\caption{{\bf Propagation dynamics of Laguerre-Gaussian wavefunctions}. (a) Probability densities of LG wavefunctions defined by indices of $\ell=1$, $p=0$ and $\ell=2$, $p=2$ upon propagation through a linear potential of strength $\alpha$. Much like the HG wavefunctions shown in Fig.~\ref{fig:HG}, the propagation of the LG wavefunctions is parametrized by the variables $w_0$ and $p_0$. Their initial profile in the $z=0$ plane are provided in the insets of the plots. (b) Transverse phase profile of an $\ell=1$ and $p=0$ LG wavefunction at various propagation distances. (c) Current lines of an LG wavefunction defined by indices $\ell=1$ and $p=0$ in the region where its probability density is maximal. Cases attributed
to various potential strengths $\alpha$ are considered.}
	\label{fig:LG}
\end{figure}

As implied by Eqs.~(\ref{eq:wavefunctionX},\ref{eq:wavefunctionY}), mode dependent features remain unaltered by the linear potential. Such features include the Gouy phase for instance, which consists of the phase components of Eqs.~(\ref{eq:wavefunctionX},\ref{eq:wavefunctionY}) that depend on the mode indices $m$ and $n$. Therefore, this potential is not expected to affect the outcome of experiments that rely on the free-space propagation features of higher-order Gaussian modes and superpositions of the latter. To illustrate this concept, we have plotted the probability density functions of Laguerre-Gaussian (LG) wavefunctions, which can be expressed as a superposition of HG wavefunctions. These modes are denoted by two integer indices $\ell$ and $p$ such that $p\geq0$. The probability densities of modes defined by indices of $\ell=2$, $p=0$ and $\ell=2$, $p=2$ upon propagation can be found in Fig.~\ref{fig:LG}(a).

\begin{figure*}[t]
	\centering
	\includegraphics[width=\linewidth]{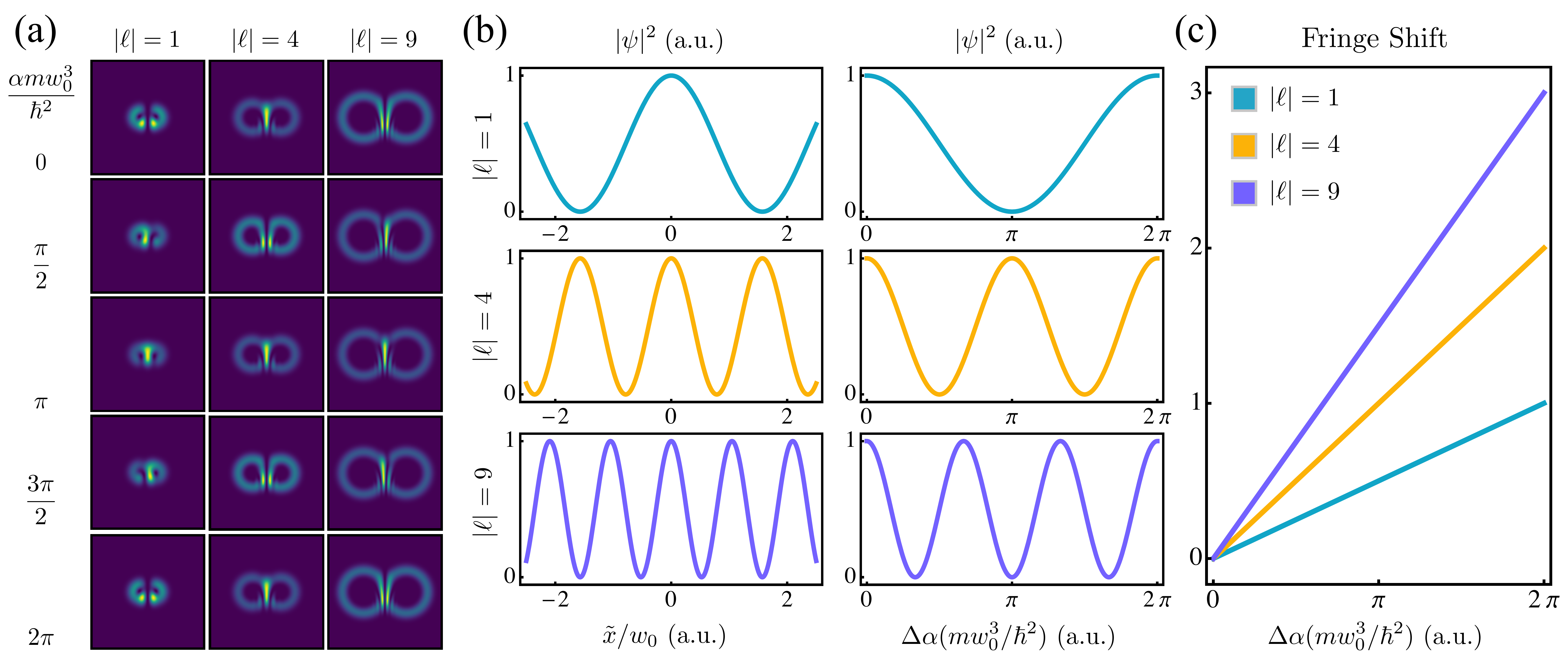}
	\caption{{\bf Vortex-based interferometry in a linear potential}. (a) Pairs of LG modes of opposite $\ell$ values are initially separated at the $z=0$ plane. After propagating for a certain distance, $z_R$ in the illustrated case, they begin to overlap and the curvature of their wavefronts causes the formation of fringes. These fringes can thereafter be shifted upon experiencing potentials defined by increasing values of $\alpha$. (b) Fringe patterns formed by the matter wave's probability density $|\psi|^2$ attributed to the interference shown in (a) in the vicinity of the crossing of the two beams centered at $\tilde{x}=0$. The left panel displays the fringe pattern formed in real space along the $\tilde{x}$ axis while the right panel shows the shifting fringe pattern associated with variations in the strength $\alpha$ of the linear potential. (c) Amount of fringe shifts observed in the beams' crossing as the strength of the linear potential is increased.}
	\label{fig:vortexInterference}
\end{figure*}

LG modes having a non-zero $\ell$ index are defined by an azimuthally-dependent phase accounted by an $\exp{(i\ell \varphi)}$ component in their wavefunction, thus causing their wavefronts to consist of $\lvert\ell \rvert$ intertwined helices with a handedness determined by the sign of $\ell$. This feature causes such wavefunctions to be eigenstates of the $z$ component of the orbital angular momentum operator, $\hat{L}_z=-i\hbar\,\partial_z$, and to be defined by $\hbar \ell$ units of OAM per particle. In addition, the influence of these helical wavefronts is manifested within the internal structure of the wavefunction itself. Namely, the presence of a phase singularity along the beam's center causes its probability density to vanish in this region. The evolution of this phase profile upon propagation is illustrated in Fig.~\ref{fig:LG}(b). As suggested by Eqs.
~(\ref{eq:wavefunctionX},\ref{eq:wavefunctionY}), the wavefunction's mode-dependent features, such as the phase singularity in this case, are preserved upon propagation, in spite of the presence of the $T^3$ phase and of the wave's tilted curvature. More interestingly, however, is the influence of the $\ell$ index on the wavefunction's probability current density $\mathbf{j}= -i\hbar \left(\psi^* \boldsymbol{\nabla} \psi - \psi \boldsymbol{\nabla} \psi^* \right)/2m$. Namely, for beams where $p=0$, the wavefunction's current lines, where the probability density is at its maximum, form skewed trajectories upon propagation~\cite{Berry08,Bli17}. These trajectories are often attributed to the classical trajectories of the particles forming the beam. In Fig.~\ref{fig:LG}(c), we illustrate how these trajectories are modified for differing values of the potential strength $\alpha$ in the case of an $\ell$=1, $p=0$ wavefunction.
Upon increasing the latter, we observed that these trajectories simply become parabolically bent around the centroid of the wavefunction's probability density.

The bending of these current lines seemingly implies that it could also change the $z$ component of the wavefunction's OAM. However, a quick calculation of the expectation value of this quantity, i.e. $\bra{\psi_\ell}\hat{L}_z\ket{\psi_\ell}$, at a given $z$ plane where the origin of the $x$ axis is shifted to the wavefunction's center of mass, reveals that it remains fixed at $\hbar\ell$. The expectation values of the other components of the OAM operator can also be calculated in a similar fashion, thus yielding values of $\bra{\psi_\ell}\hat{L}_x\ket{\psi_\ell}=0$ and $\bra{\psi_\ell}\hat{L}_y\ket{\psi_\ell}=-z^2 \alpha m/p_0$. Note that the second value simply consists of the OAM attributed to the parabolic trajectory of the particle in the $xz$ plane, and therefore corresponds to an extrinsic form of OAM that cannot be attributed to the internal structure of the wavefunction itself.

The mode invariance of the propagation features related to the potential's presence could be of use in new interferometric schemes less traditional than the ones shown in Fig.~\ref{fig:interferometer}. For instance, one could use the increased transverse extent of LG wavefunctions, which allows them to sample more of their curvature, for such purposes. The principles of such a scheme are depicted in Fig.~\ref{fig:vortexInterference}, where two initially separated LG beams with opposite values of $\ell$, are made to propagate under the influence of the potential. After a certain distance, the edge of these beams, which roughly have the same phase, overlap, thus forming an interference pattern due to the presence of the curvature of their wavefronts shown in Fig.~\ref{fig:vortexInterference}(a). Pairs of beams with higher values of $|\ell|$ extend over regions defined by a larger curvature, thereby producing thinner fringes in which small variations of $\alpha$ are more observable. Upon experiencing an increase in this potential, the fringes of the pattern will shift, thus providing information pertaining to the potential without relying on interferometric schemes that use diffraction gratings. Both of the spatial fringes in the proximity of the beams' crossings and their variations with the strength of the potential scale with $\sqrt{|\ell|}$ as depicted in Fig.~\ref{fig:vortexInterference}(b). This causes more fringe shifts to occur upon varying the $\alpha$ parameter of the potential, thereby making beams carrying larger values of OAM more sensitive to perturbations in $\alpha$. This increased sensitivity is illustrated in Fig.~\ref{fig:vortexInterference}(c).

\section{Conclusion}

In conclusion, we have derived Gaussian paraxial solutions describing the propagation of structured matter waves within a linear potential. We demonstrate that effects related to the presence of the potential globally affect the structure of the propagating wavefunction without displaying any mode-dependent features. We also provide eigenvalue decompositions of these Gaussian solutions and relate them to their Airy transforms. Finally, we apply our analysis to the dynamics of matter waves carrying phase vortices by analyzing their wavefunction's spatial profile upon propagation and their corresponding probability density current lines. We also suggest that the symmetry of such solutions could be of interest in interferometry experiments that aim to measure the strength of linear potentials.

The derived formalism could be useful in further investigations addressing the propagation of structured matter waves in linear potentials. For instance, the derived wavefunctions could be employed to analyze the dynamics of longitudinally structured waves. Furthermore, the derived expansion coefficients could provide a means for analyzing the influence of additional perturbative potentials on the deflected waves while also being of mathematical interest in the calculation of Airy transforms of functions corresponding to the wavefunctions of propagating paraxial structured waves.

\section{acknowledgement}
This work was supported by Ontario's Early Researcher Award (ERA), Canada Research Chairs (CRC), and the European Union's Horizon 2020 Research and Innovation Programme (Q-SORT), grant number 766970. R.F. acknowledges the financial support of the Banting postdoctoral fellowship of the NSERC.

\bibliography{references}

\end{document}